\documentclass[%
 reprint,
 superscriptaddress,
 amsmath,amssymb,
 aps,
 pra,
 floatfix,
]{revtex4-2}

\usepackage{graphicx}
\usepackage{dcolumn}
\usepackage{bm}
\usepackage{hyperref,bookmark}
\hypersetup{hidelinks}

\usepackage{CJK} 
\usepackage{color}
\usepackage{ulem}
\usepackage{braket}
\usepackage[caption=false]{subfig}
\usepackage[title]{appendix}

\begin{document}

\begin{CJK*}{UTF8}{gbsn}  
%%%%%%%%%%%%%%%%%%%%%%%%%%%%%%%%%%%%%%%
\title{Post-selection shifts the transition frequency of helium in an atomic beam}

\author{Jin-Lu Wen (温金录)$^{a,}$ }
\affiliation{Department of Chemical Physics, University of Science and Technology of China, Hefei 230026, China}

\author{Jia-Dong Tang (唐家栋)$^{a,}$}
\affiliation{Hefei National Research Center of Physical Sciences at the Microscale, University of Science and Technology of China, Hefei 230026, China}

\author{Ya-Nan Lv (吕亚男)$^{a,}$}
\affiliation{Hefei National Research Center of Physical Sciences at the Microscale, University of Science and Technology of China, Hefei 230026, China}

\author{Yu R. Sun (孙羽)}
\email{sunyu@mail.iasf.ac.cn}
\affiliation{Institute of Advanced Science Facilities, Shenzhen, 518107, China}

\author{Chang-Ling Zou (邹长铃)}
\email{clzou321@ustc.edu.cn}
\affiliation{CAS Key Laboratory of Quantum Information, University of Science and Technology of China, Hefei 230026, China}

\author{Jun-Feng Dong (董俊峰)}
\affiliation{Hefei National Laboratory, University of Science and Technology of China, Hefei 230088, China}

\author{Shui-Ming Hu (胡水明)}
 \email{smhu@ustc.edu.cn}
\affiliation{Department of Chemical Physics, University of Science and Technology of China, Hefei 230026, China}
\affiliation{Hefei National Research Center of Physical Sciences at the Microscale, University of Science and Technology of China, Hefei 230026, China}
\affiliation{Hefei National Laboratory, University of Science and Technology of China, Hefei 230088, China}

\date{\today}

\begin{abstract}

Post-selecting output states in measurements can effectively amplify weak signals and improve precision. However, post-selection effects may also introduce unintended biases in precision measurements. Here, we investigate the influence of post-selection in the precision spectroscopy of the $2^3S - 2^3P$ transition of helium ($^4$He) using an atomic beam. We directly observe that post-selection based on atomic positions causes a shift in the measured transition frequency, amounting to approximately -55 kHz. After accounting for this post-selection shift, we obtain a corrected frequency of $276,764,094,712.45 \pm 0.86$ kHz for the $2^3S_1 - 2^3P_0$ transition. Combining this result with existing data for $^3$He, we derive a new value for the difference in squared nuclear charge radii, $\delta r^2 [r_{h}^{2} - r_{\alpha}^{2}] = 1.0733 \pm 0.0021$ fm$^2$. This value shows a $2.8\sigma$ deviation from measurements of muonic helium ions, potentially pointing to new physics that challenges lepton universality in quantum electrodynamics.

\end{abstract}

\maketitle

%%%%%%%%%%%%%%%%%%%%%%%%%%%%%%%%%%%%%%%%%%%%%
\section{Introduction}\label{sec:Int}
%%%%%%%%%%%%%%%%%%%%%%%%%%%%%%%%%%%%%%%%%%%%%

Over the past few decades, quantum mechanics has profoundly reshaped our understanding of nature, facilitating the development of technologies that defy classical intuitions.
One of the intriguing quantum effects is post-selection (PS) and weak value amplification (WVA)~\cite{Aharonov1988PRL_wva, Dressel2014, Aharonov2010}. 
By selectively detecting outcomes in a weak interaction measurement, PS and WVA can enhance the measurement signal, though this comes at the cost of discarding part of the observed data~\cite{Denkmayr2014}. Despite some controversy surrounding its counterintuitive nature in recent years~\cite{Vaidman2017}, WVA has found applications in precision measurement~\cite{Hosten2008} and inspired new ideas in quantum technologies~\cite{Kocsis2011, Goggin2011, Lundeen2011, Feizpour2011, Jordan2014, Hallaji2017, Liu2020NC}.
This quantum mechanical effect has particular relevance for precision measurements, where even very tiny spectroscopic shifts in atoms and molecules can be crucial for tasks such as maintaining accurate time~\cite{Markowitz1958PRL_time},
detecting weak forces~\cite{Safronova2018PRL_newphysics, Ficek2017PRA_5thforce}, and testing fundamental physical models~\cite{Adkins2022PR_Ps, Figueroa2022PRL_YbIS}.
Although post-selection can enhance the precision of measurements involving multiple degrees of freedom, it may also introduce systematic shifts that must be carefully accounted for. 

In this work, we investigate the impact of post-selection on the precision spectroscopy of the $2^3 S-2^3 P$ transition of $^4$He in an atomic beam. We observed an unexpected discrepancy in the results depending on how the experimental data is analyzed, which we attribute to PS effects. This interpretation is supported by theoretical analysis and simulations. Our findings underscore the importance of carefully considering subtle quantum effects, like post-selection, in high-precision measurements.

Precise measurements of helium atom transition frequencies have long been used to determine nuclear charge radii, as electron penetration into the non-pointlike nucleus causes energy level shifts that depend on the nucleus's size. 
Recent spectroscopy~\cite{Werf2023arxiv_he3_trap, Rengelink2018_He4_trap, Rooij2011Science_He4_23S-21S} of the $2^3S-2^1S$ transitions in $^3$He and $^4$He provided the difference between the squared charge radii of these two nuclei, $\delta r^2 [r_{h}^{2} - r_{\alpha}^{2}] = r^2$($^3$He) $- r^2$($^4$He). This result deviated by $3.6\sigma$ from measurements derived from spectroscopy~\cite{CREMA2023arxiv_uhe3, Krauth2021Nature_uHe4} of the muonic helium ion ($\mu$-He$^+$), in which the electron is replaced by a muon.
A similar discrepancy was noted in measurements of the proton charge radius, which has been studied extensively using hydrogen atom ($e$-H) spectroscopy~\cite{Matveev2013PRL_H_Lamb}, but shows a significant deviation when measured using muonic hydrogen ($\mu$-H)~\cite{Antognini2010Science_uH}, leading to the well-known ``proton radius puzzle''. 
In light of these findings, our results of $^4$He, combined with the $^3$He result of Cancio Pastor \textit{et al.}~\cite{Pastor2012PRL_He3},  allow us to compare the $\delta r^2$ value from $e$-He spectroscopy with that from the precise spectroscopy of $\mu$-He$^+$, providing a rigorous test of the equivalence of leptons in electromagnetic interactions. 
Should these deviations be confirmed, they could pose a significant challenge to the Standard Model, as the muon and electron are believed to share identical electromagnetic properties in quantum electrodynamics (QED).

%%%%%%%%%%%%%%%%%%%%%%%%%%%%%%%%%%%%%%%%%%%%%
\section{Experimental observation of the post-selection effect}\label{sec:exp}
%%%%%%%%%%%%%%%%%%%%%%%%%%%%%%%%%%%%%%%%%%%%%

%%%%%% Fig1
\begin{figure*}[t]
\centering
\includegraphics[width=0.95\textwidth]{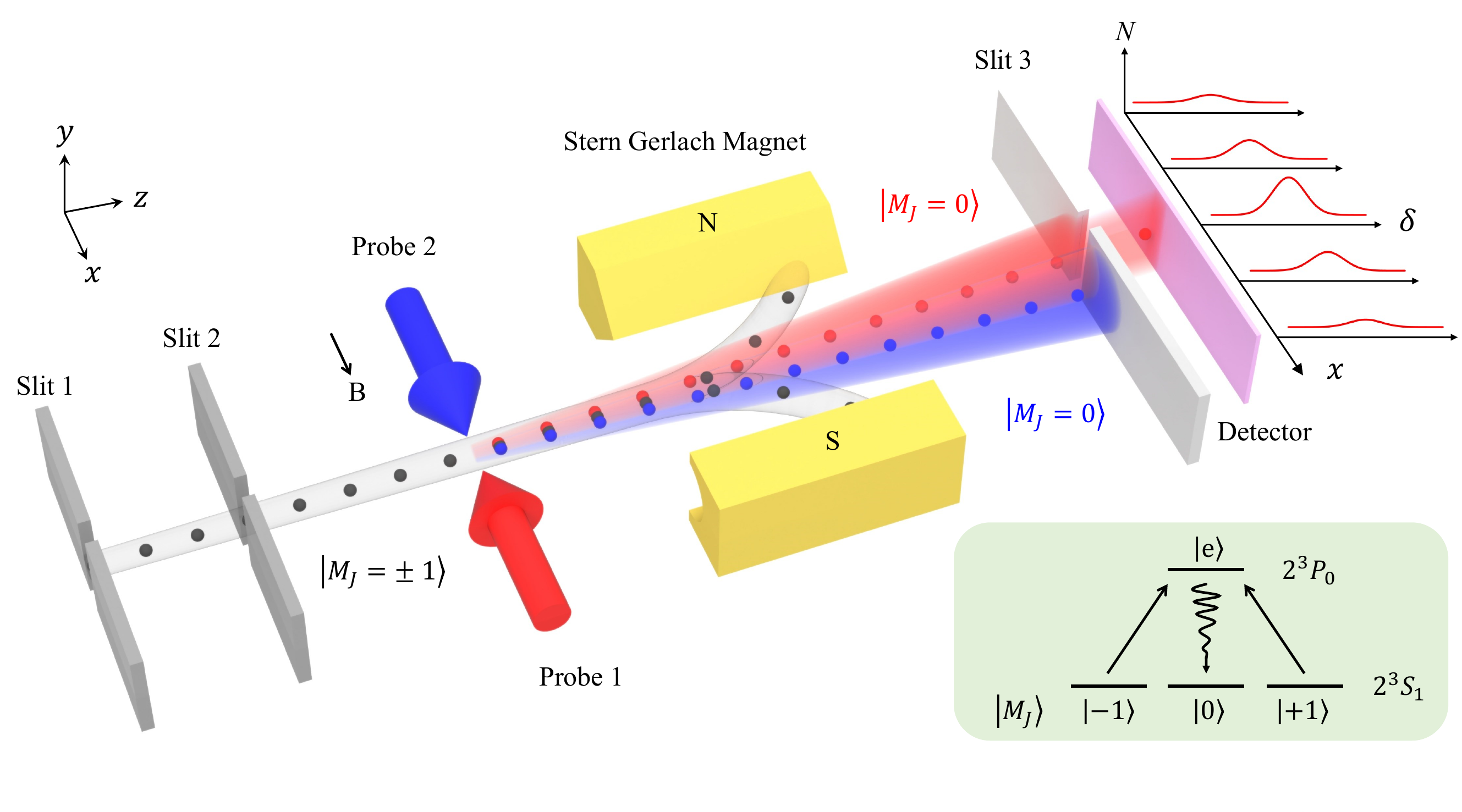}
\caption{\textbf{Experimental Setup of precision spectroscopy of atomic helium with post-selection.} 
A collimated helium atomic beam is prepared in the $M_J = \pm 1$ levels of the $2^3 S_1$ state (lifetime $\tau \sim 7900$~s). 
The probe laser is scanned around the resonance of the $2^3 S_1 - 2^3P_0$ transition.
Two counter-propagating probe laser beams, labeled ``Probe 1'' and ``Probe 2'', are periodically blocked, so that at any given time, only one beam interacts with the atoms.
Atoms excited by the probe laser to the $2^3 P_1, M_J=0$ state undergo spontaneous decay ($\tau \sim 98$~ns) to either the $2^3 S_1, M_J=0$ or $2^3 S_1, M_J=\pm 1$ levels. 
Atoms in the $M_J=\pm 1$ states are deflected by a Stern-Glarch magnet (SGM) and removed from the atomic beam, while only atoms in the $M_J=0$ state pass through a narrow slit (width $\Delta x_1 \sim 0.3$~mm) placed before a multi-channel plate (MCP) detector. These atoms, shown as blue and red spheres, correspond to atoms interacting with Probe 1 and Probe 2, respectively.
The slit can be moved along the $x$ axis to select atoms with specific momenta. 
As atoms absorb and emit photons during the interaction, their velocity in the $x$ direction changes, affecting the position where they reach the slit.  
For a given $x$ position of the slit, spectra are recorded by counting the number of detected atoms ($N$) while scanning the laser frequency ($\delta$). The central frequencies obtained from both probe beams are averaged to eliminate the first-order Doppler shift.
}
\label{fig:setup}
\end{figure*}
%%%%%%

We employed an atomic beam to measure the  $2^3 S - 2^3 P$ transition of helium and determine the nuclear charge radii difference between $^3$He and $^4$He~\cite{Shiner1995PRL_He3, Pastor2012PRL_He3, Rengelink2018_He4_trap, Rooij2011Science_He4_23S-21S, Huang2020pra_2s3d, Sick2014PRC_zemach}.
Fig.~\ref{fig:setup} shows the experimental setup, where a collimated beam of helium atoms interacts with probe lasers. 
Atoms with specific internal and external states pass through a slit (slit3 in Fig.~\ref{fig:setup}) before reaching the detector. This method, first proposed by Ramsey~\cite{ramsey1949pr_beammethod}, has been widely adopted in precision measurements due to its simplicity and ability to explore systematic uncertainties. 
In our experiment, we applied the SCTOP (sequential counter-propagating traveling-wave optical probing) method~\cite{Wen2023PRA_traveling}, which involves three key steps:
(i) State preparation: Helium atoms are prepared in the $| M_J =\pm1 \rangle$ levels of the $2^3 S_1$ metastable state, the lateral momentum of the atoms is denoted as $p = m v_x$, while the longitudinal velocity $v_z$ can be adjusted within the range of $100 - 400$~m/s with a spread of about $\pm 5$~m/s~\cite{Chen2020PRA_He}. 
(ii) Interrogation: The atomic transition is probed by a tunable $1083\,\mathrm{nm}$ laser. When the laser is resonant with the transition to the $2^3 P_0$ state, atoms are probably to absorb photons and spontaneously decay (with a natural lifetime of $98$~ns) to the $| M_J=0\rangle$ ground state.
To eliminate the first-order Doppler shift caused by atoms with non-zero lateral velocity $v_x$, two counter-propagating laser beams are used~\cite{Wen2023PRA_traveling}.
(iii) Detection: Only the atoms in the $| M_J=0 \rangle$ state pass through a Stern-Gerlach magnet without deflection and are collected by a microchannel plate (MCP) detector for counting.

%%%%%%%
\begin{figure*}[ht]
\centering
\subfloat{\label{fig:2a}}
\subfloat{\label{fig:2b}}
\subfloat{\label{fig:2c}}
\subfloat{\label{fig:2d}}
\subfloat{\label{fig:2e}}
\subfloat{\label{fig:2f}}
\includegraphics[width=0.95\textwidth]{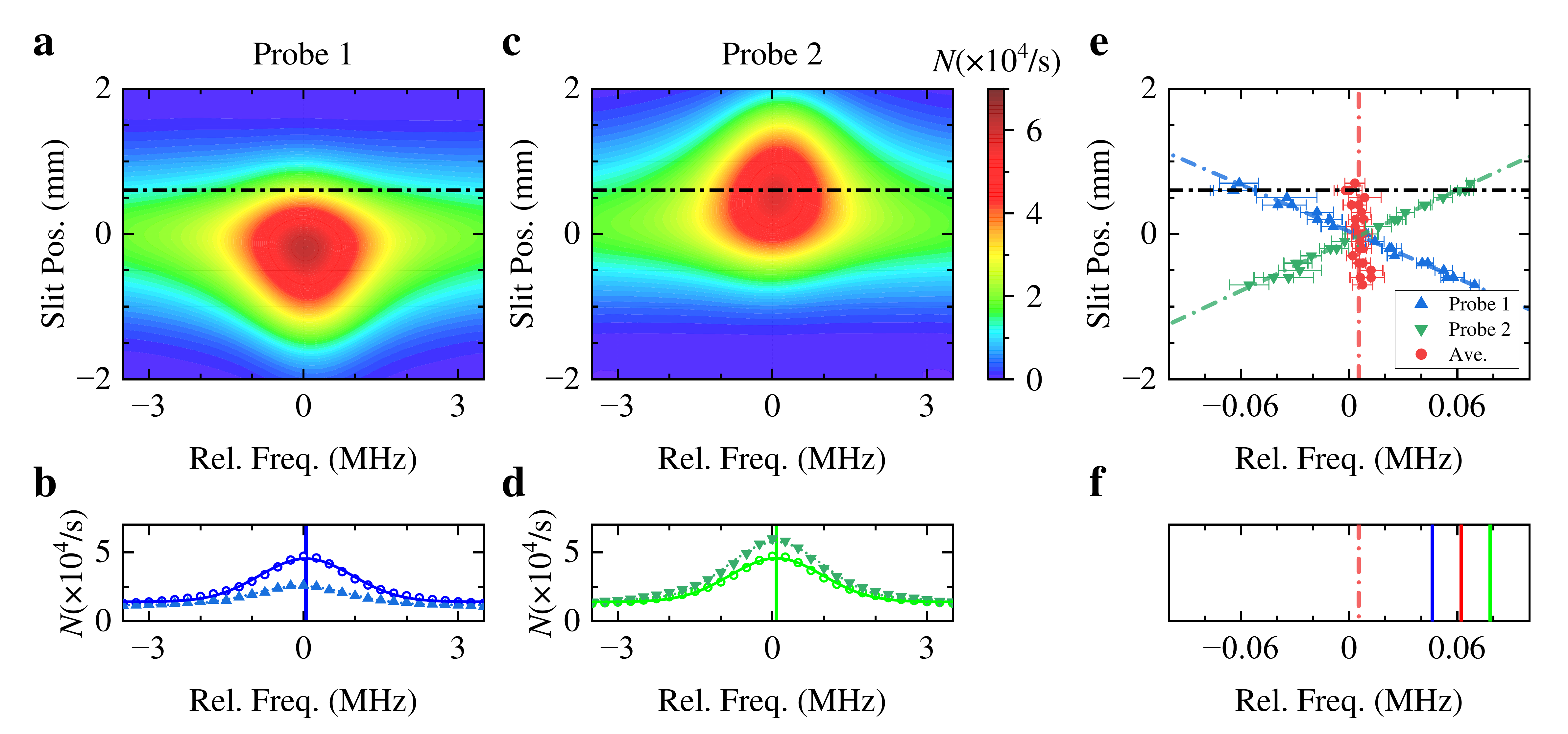}
\caption{\textbf{Experimental data of the spectroscopy of the $\boldsymbol{2^3S-2^3P}$ transition in a $\boldsymbol{^4}$He atomic beam probed by the SCTOP method.}
\textbf{a}, Experimental data recorded by Probe 1 at different slit positions along the $x$ axis. 
\textbf{b}, Solid triangles represent the spectrum obtained with post-selection of atoms at the position $x=+0.6$~mm (dashed-dotted line on \textbf{a}), while open circles show the spectrum without post-selection. 
\textbf{c} \& \textbf{d}, The corresponding results obtained with Probe 2. 
\textbf{e}, Center frequencies of the spectra with post-selection at each $x$ position for Probe 1 (blue up-triangles) and Probe 2 (green down-triangles). Averaged center frequencies at various $x$ positions are shown as red solid circles. 
\textbf{f}, Center frequencies obtained with post-selection (red dotted line) compared to those without post-selection (solid lines). The blue and green lines correspond to center frequencies without post-selection for Probe 1 and Probe 2, respectively, and their average is represented by the red solid line.
}
\label{fig:dist}
\end{figure*}
%%%%%

During the interrogation, changes in the internal atomic state are linked to changes in external degrees of freedom. 
After absorbing and emitting photons, the lateral momentum of a helium atom changes from $p=mv_x$ to $p=mv_x+\hbar k - \hbar r$, where $\hbar k$ is the momentum of the probe laser photon and $\hbar r$ represents the randomly emitted photon. 
Repeated absorption and emission cycles can cause significant momentum shifts as the atom transitions between internal states ($|M_J = \pm1\rangle$ to $|M_J = 0\rangle$), creating an entanglement between momentum change and atomic state transitions. This entanglement allows for weak value amplification (WVA) of atomic momentum, as schematically illustrated in Fig.~\ref{fig:setup}. 
By selecting atoms based on their position $x$ (i.e., significantly away from the center), we can observe large momentum shifts associated with considerable changes in the resonance frequency, together with a reduced number of detected atoms. 
Measurements of the amplified Doppler shift enable a precise evaluation of the systematic shift, facilitating the determination of the transition frequency. 
We placed a narrow slit in front of the detector to select atoms with specific momenta based on the location of the narrow slit ($x$). Additionally, the two counter-propagating probes (``Probe 1'' and ``Probe 2'') were alternatively blocked~\cite{Wen2023PRA_traveling}, and two spectra were recorded at each $x$. A feedback servo system was used to ensure the angle deviation between the two beam directions remains below $1\times 10^{-5}$~rad. 

Fig.~\ref{fig:2a} illustrates the relationship between the detected number of atoms and their position $x$ and the frequency of Probe 1, with an atomic velocity of $v_z=290$~{m/s}. Most detected atoms are located near the origin $x\approx 0$ but show a slight shift along the $x$-direction due to the probe laser. 
We obtained a spectrum at each $x$. For instance, the spectrum indicated by solid triangles in Fig.~\ref{fig:2b} corresponds to the location $x=0.6$~mm marked by the dashed-dotted line in Fig.~\ref{fig:2a}. 
Each spectrum was fitted using a Lorentzian function, and the fitted center, along with its uncertainty, is represented as a blue triangle with an error bar in Fig.~\ref{fig:2e}.
The observed line width is approximately 2.2~MHz,which aligns with the natural linewidth of the $2^3S-2^3P$ transition (1.6~MHz). 
When only the counter-propagating probe laser (Probe 2) was used, results are depicted in Figs.~\ref{fig:2c} \&~\ref{fig:2d}, with the fitted spectral centers shown as green triangles in Fig.~\ref{fig:2e}. Notably, signals weaken significantly at larger $x$ positions, leading us to include only results for $|x|< 0.7$~mm in Fig.~\ref{fig:2e}. 
A linear fit of the blue and green dotted lines indicates that the spectral center varies linearly with $x$,  suggesting a Doppler shift as the probe interacts with atoms. The fitted slope is approximately 0.09~MHz/mm, consistent with the calculated Doppler shift at $v_d=v_z(x/L)$, where $L$ is the distance from the probe laser to the slit.
Comparing results from the two probing directions reveals that the shift averages out for each pair of centers obtained at the same slit position. This yields a result that is independent of $x$, as represented by the red dotted lines in Figs.~\ref{fig:2e} \&~\ref{fig:2f}. 

To assess the effect of WVA, we contrast the results obtained with and without post-selection of final states. 
By averaging the counts over $x$ for all data in Figs.~\ref{fig:2a} \&~\ref{fig:2c}, we derive two spectra, shown as open circles in Figs.~\ref{fig:2b} \&~\ref{fig:2d}. 
Importantly, removing the slit in front of the detector yields identical spectra.
Both spectra were fitted with Lorentzian functions, with their centers illustrated as blue and green lines in Fig.~\ref{fig:2f}. The mean value of these centers is marked by the red solid line. A notable deviation of about $-55$~kHz is observed between the results with and without post-selection, despite utilizing the same experimental data. Measurement conducted without the slit also yielded results consistent with this value, albeit without the post-selection effect. 

%%%%%%%%%%%%%%%%%%%%%%%%%%%%%%%%%%%%%%%%%
\section{Theoretical interpretation}
%%%%%%%%%%%%%%%%%%%%%%%%%%%%%%%%%%%%%%%%%

The expected number of collected atoms through slit3 could be written as $N(\delta)\propto|\langle{\psi_f(p)}|U_{SE}\langle{e}|O(\delta)|{\pm1}\rangle|{\psi_i(p)}\rangle|^2$, where $\psi_{i/f}(p)$ represents the initial/final momentum wavefunctions of the atom, $O(\delta)$ describes the weak transition operator, flipping the internal state of the atom while transferring momentum between the photon and atom, with detuning $\delta$, and $U_{SE}$ denotes the evolution operator for spontaneous radiation from the excited state $|e\rangle$. 
To simplify the model and avoid directly solving the operator $O$, we consider the velocity distributions of the atoms in the lateral ($x$) direction. The resulting number of collected atoms is approximately:
\begin{equation}
    N(\delta)\propto\int \int dv_0dv_1\frac{P_0(v_0)P_t(v_0,v_1)P_1(v_1)}{(\delta-kv_0)^2+\gamma^2},
    \label{eq:dist}
\end{equation}
where: $P_0(v_0)=e^{-(v_0-v_{0,b})^2/2\sigma_0^2}$ represents the velocity distributions of initial atoms with a bias velocity $v_{0,b}$ and spread $\sigma_{0}$;
$P_1(v_1)=e^{-(v_1-\xi x)^2/2\sigma_1^2}$ is the velocity distribution of the detected atoms passing through the slit, where $\xi x$ is the bias velocity at detection, $\xi$ depends on the time of flight, and $\sigma_{1}$ is the velocity spread at detection;  
$P_t(v_0,v_1)=\frac {1}{2}\sqrt{1-(v_1-v_0-v_R)^2/v_R^2}$ for $0\leq \frac{1}{2}|v_1-v_0|\leq v_R$ is the probability distribution for lateral velocity change from $v_0$ to $v_1$ due to absorption and spontaneous emission of photons, with $v_R=\hbar k/m$. 
Since in practice $v_R\ll\sigma_{0,1}$, we can approximate the number of atoms as $N(\delta)\propto\int dv_0\frac{P_0(v_0)P_1(v_0+v_R)}{(\delta-kv_0)^2+\gamma^2}$. 
In the case of a narrow slit in front of the detector ($\sigma_1\ll\sigma_0$), the center of the spectrum for Probe 1 can be approximated as $\delta_{\mathrm{PS},+}=k(\xi x-v_R)$, 
and for Probe 2, the center is $\delta_{\mathrm{PS},-}=-k(\xi x + v_R)$. 
The post-selection can indeed produce a considerable Doppler shift in the measured spectral center, which is linearly dependent on the slit position $x$. This observation aligns with the results presented in Fig.~\ref{fig:2e}.
For measurements without post-selection, where the detector slit is wide enough that $\sigma_1\gg \gamma/k$, the atomic velocity distribution $P_1(v_0+v_R)$ can be approximated as constant. In this case, the spectral centers are primarily determined by the initial distribution of the atomic velocity, yielding shifts of $\delta_{\mathrm{NPS},+}\approx kv_{0,b}$ for Probe 1 and $\delta_{\mathrm{NPS},-}\approx -kv_{0,b}$ for Probe 2.
Notably, the $v_{0,b}$-dependent frequency shift can be mitigated by averaging the measured spectral centers obtained with counter-propagating probe lasers.
However, for the post-selection scenario, a residual frequency shift of $(\delta_{\mathrm{PS},+}+\delta_{\mathrm{PS},-})/2=-k v_R$ persists when compared to results without post-selection. This shift is attributed to the recoil of the atoms upon photon absorption, as illustrated in Fig.~\ref{fig:2f}. 

%%%%%%%%%%%%
\begin{figure}[ht]
\centering
\includegraphics[width=0.45\textwidth]{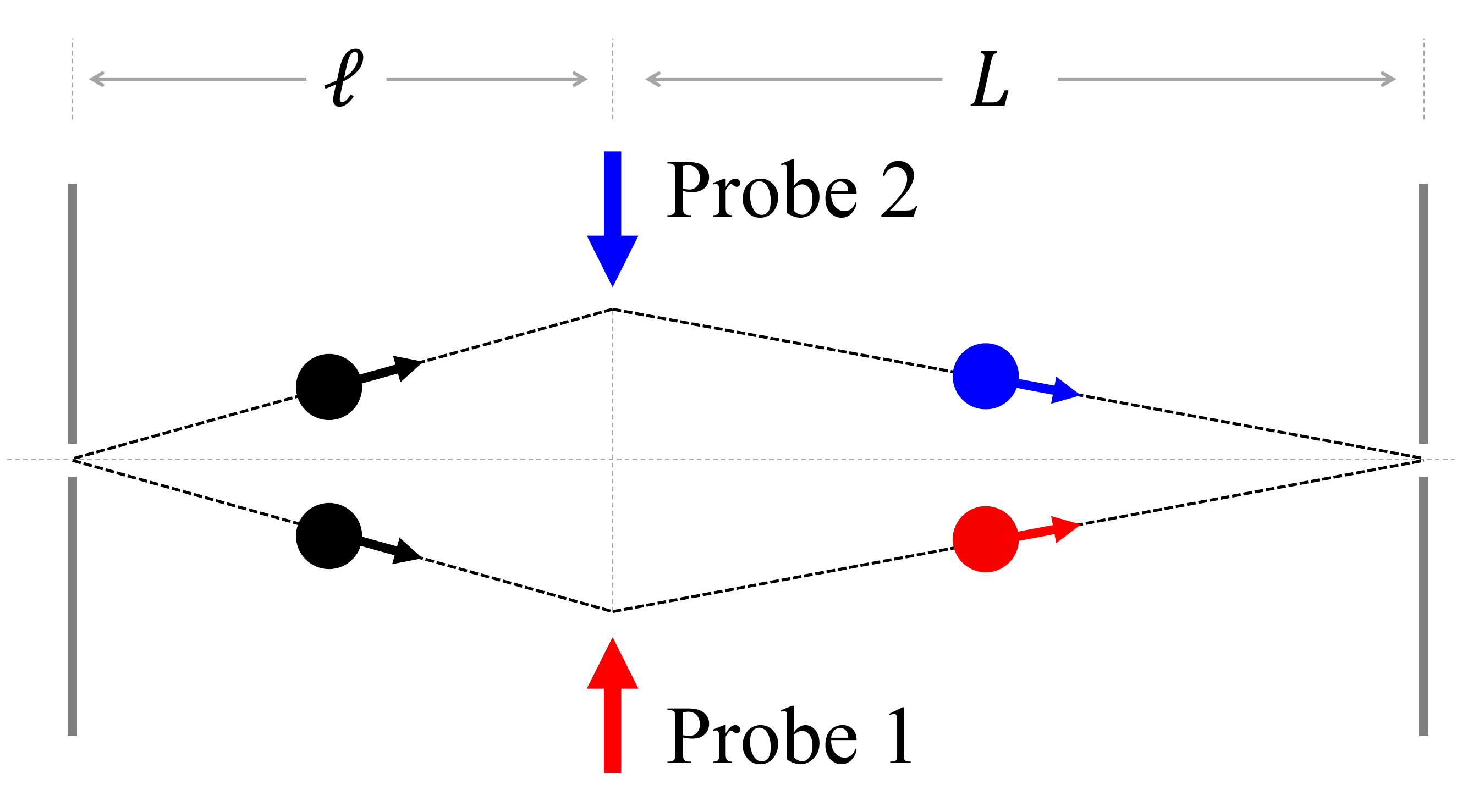}
\caption{\textbf{A phenomenological model of the post-selection effect.} Two slits were placed to select atoms with a zero velocity along the x-axis. However, due to the momentum change after absorbing a photon from the probing laser, only atoms with a nonzero initial velocity $v_x \simeq v_R\frac{L}{\ell+L}$ can pass the second slit, which induces a red shift in the center frequency. Note that the same shift applies with a counter-propagating probe beam (Probe 2).}
\label{fig:pheno}
\end{figure}
%%%%%%%%%%%%

The experimental results can also be understood through a phenomenological model (see Appendix~\ref{app:phenome}), as depicted in Fig.~\ref{fig:pheno}. Two slits are used to select atoms with zero velocity along the axis of the probing laser beams. 
When an atom absorbs a photon from the Probe 1 laser, its momentum changes. Only atoms with a specific negative initial velocity of $-v_R\frac{L}{\ell+L}$ along the $x$ direction, counter-propagating to the Probe 1 beam, can pass the second slit and be detected, where $v_R$ is the recoil velocity from photon absorption, $\ell$ is the distance from the atom-laser interaction point to the first slit, and $L$ is the distance to the second slit. 
In this case, the first-order Doppler effect induces a red-shift ($\Delta \nu$) to the transition frequency ($\nu_c$): 
\begin{equation}\label{eq:vx}
    \frac{\Delta \nu}{\nu_c} = - \frac{v_R}{c}\frac{L}{\ell+L}.
\end{equation}
When Probe 2 is active, only atoms with a reversed velocity can pass the final slit. However, since Probe 2 propagates in the opposite direction, the center frequency is red-shifted again.
This phenomenon illustrates how both probe beams in opposite directions reinforce the same red shift in the detected spectra.

Both the experimental results (Fig.~\ref{fig:2f}) and the theoretical analysis (detailed in Appendix ~\ref{app:theory}) reveal a systematic frequency shift due to the post-selection effect, referred to as the post-selection shift (PSS), which was measured to be $\Omega_{\mathrm{PSS}}\approx$ -55~kHz. 
This shift, described by Eq.~\ref{eq:vx}, occurs when a very narrow slit3 is used. 
PSS is expected to vanish if slit3 is positioned close to the probing zone or made sufficiently wide to detect all atoms, regardless of their velocity along the x-axis. 

To mitigate this, we utilized a narrow incident atomic beam~\cite{Wen2023PRA_traveling} and a wide slit3 (about 10~mm) to measure the $2^3 S_1-2^3 P_0$ transition. With the PSS effect effectively eliminated, we determined the transition frequency of $^4$He to be 276\,764\,094\,712.45 $\pm$ 0.86~kHz. By combining these measurements with the fine structure intervals reported in a previous study~\cite{Zheng2017PRL_FS}, we calculated the centroid frequency of the $2^3 S-2 ^3 P$ transition to be 276\,736\,495\,655.21 $\pm$ 0.87~kHz. 
Furthermore, the PSS can be derived and corrected in certain experimental configurations, as outlined in the Appendix~\ref{app:pss}. 

%%%%%%%%%%%%%%%%%%%%%%%%%%%%%%%%%%%%%%%%%
\section{Nuclear charge radius square difference between \texorpdfstring{$^3$}{3}He and \texorpdfstring{$^4$}{4}He } \label{Sec:3He-4He}
%%%%%%%%%%%%%%%%%%%%%%%%%%%%%%%%%%%%%%%%%

%%%%%%%%%%%%
\begin{figure}[ht]
\centering
\includegraphics[width=0.45\textwidth]{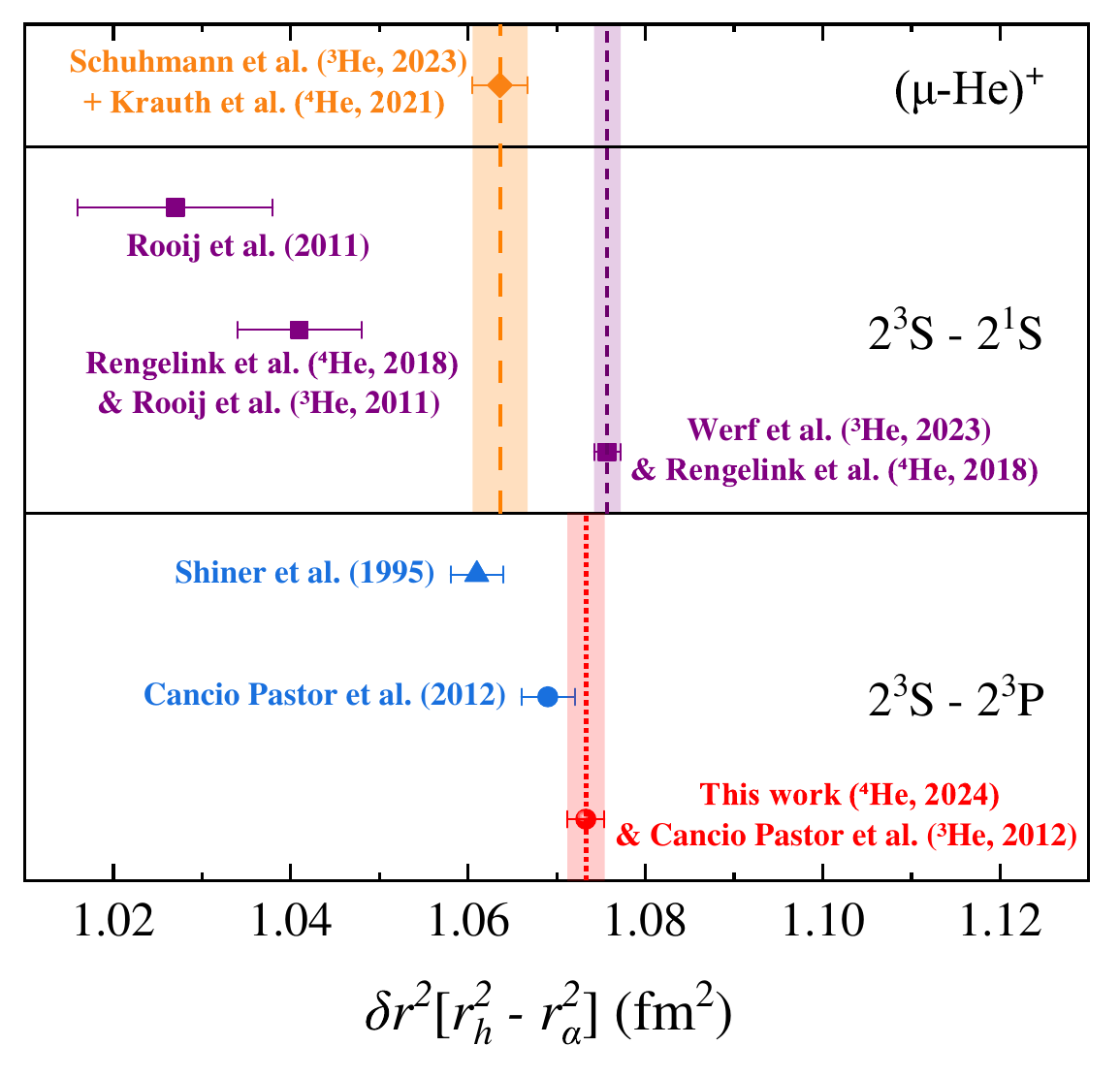}
\caption{\textbf{The difference between the squared nuclear charge radii of $\boldsymbol{^3}$He and $\boldsymbol{^4}$He.} 
The results are derived from isotope shifts in different transitions. The orange circle represents the result for the muonic helium ion~\cite{Krauth2021Nature_uHe4,CREMA2023arxiv_uhe3}. The purple squares represent the results of the $2^3 S - 2^1 S$ transition obtained by the Amsterdam group~\cite{Rooij2011Science_He4_23S-21S,Rengelink2018_He4_trap,Werf2023arxiv_he3_trap}. The previous result has recently been updated to the new one shown on the right. The bottom panel shows the results obtained by three groups detecting the $2^3S - 2^3 P$ transition. 
Note that the result of Shiner's group~\cite{Shiner1995PRL_He3} (triangle) was also affected by the post-selection effect.}
\label{fig:isotope}
\end{figure}
%%%%%%%%%%%%

The precise determination of transition frequencies plays a crucial role in testing the standard model, particularly with respect to accounting for the finite size of the helium nucleus. This finite size results in frequency shifts compared to a point-like nucleus. 
In isotopic shifts between $^4$He and $^3$He atoms, most mass-independent terms cancel out, allowing for the precise determination of the difference in squared nuclear charge radii between the two isotopes, denoted as $\delta r^2 [r_{h}^{2} - r_{\alpha}^{2}] $.
By combining the $^4$He transition frequency determined in this work with the $^3$He value from Cancio Pastor \textit{et al.}~\cite{Pastor2012PRL_He3}, a new $\delta r^2$ value of 1.0733 $\pm$ 0.0021~fm$^2$ is derived. This result is shown in Fig.~\ref{fig:isotope} as a red circle, compared with results from other groups. 
Notably, Shiner's group, using a similar approach based on the $2^3S-2^3P$ transition~\cite{Shiner1994PRL_He4,Shiner1995PRL_He3}, derived $\delta r^2 $ = 1.061 $\pm$ 0.003~fm$^2$, which significantly deviates from our results and may be influenced by the PSS effect. However, since the PSS value can differ between $^3$He and $^4$He isotopes, we cannot estimate a direct PSS correction to Shiner's result. 
The $\delta r^2$ value has also been measured using the forbidden $2^3 S-2 ^1 S$ transition in cold atomic clouds~\cite{Rengelink2018_He4_trap, Rooij2011Science_He4_23S-21S}, with a revised value of $1.0757 \pm 0.0015$~fm$^2$ reported recently~\cite{Werf2023arxiv_he3_trap}.
Additionally, the latest measurements from muonic helium ions~\cite{Krauth2021Nature_uHe4,CREMA2023arxiv_uhe3} yield $\delta r^2$ = 1.0636 $\pm$ 0.0031~fm$^2$, showing an approximately $2.8\sigma$ deviation from the value obtained from electronic helium spectroscopy.

While the cause of this discrepancy remains unclear, it may be necessary to consider the potential post-selection effect in various experiments~\cite{Tiesinga2021RevModPhys_CODATA2018}, such as the helium ion experiment~\cite{vanWijingaarden2000PRA_HeionLamb}, Rydberg-state spectroscopy~\cite{Hessels1992PRA_HeRydberg}, and hydrogen atom spectroscopy~\cite{Berkeland1995PRL_H_PS}.
If these deviations between electronic and muonic measurements in both hydrogen and helium are confirmed, it could suggest a non-equivalence of leptons in electromagnetic interactions, pointing to new physics beyond the Standard Model.

%%%%%%%%%%%%%%%%%%%%%%%%%%%%%%%%%%%%%%%
\begin{acknowledgments}
This work was jointly supported by the Ministry of Science and Technology of China (Grant No. 2021ZD0303102), the National Natural Science Foundation of China (Grants No. 22241301, No. 91736101, and No. 21688102), and the Strategic Priority Research Program of the Chinese Academy of Sciences (Grants No. XDB21010400 and No. XDC07010000).
\end{acknowledgments}
%%%%%%%%%%%%%%%%%%%%%%%%%%%%%%%%%%%%%%%

%%%%%%%%%%%%%%%%%%%%%%%%%%%%%%%%%%%%%%%

\begin{appendices}

%%%%%%%%%%%%%%%%%%%%%%%%%%%%%%%%%%%%%%%%%%%%%%%%%%%%%%%%%
\section{ Measurement of the \texorpdfstring{$^4$}{4}He transition frequency and the isotope shift}~\label{app:measurement}
%%%%%%%%%%%%%%%%%%%%%%%%%%%%%%%%%%%%%%%%%%%%%%%%%%%%%%%%%

This section outlines the process for measuring the transition frequency of $^4$He from the $2^3 S_1$ to the $2^3 P_0$ state. 
The process begins by exciting helium atoms to the $2^3 S_1$ state using a radiofrequency discharge.
The atoms are then slowed using a Zeeman slower, collimated with a 2D magneto-optical trap (MOT), and subsequently deflected by a laser tuned to the $2^3 S_1 - 2^3 P_0$ transition. 
The atoms are pumped into the $2^3 P_0$ state and spontaneously decay to the $M = \pm 1$ sublevels of the $2^3 S_1$ state. A Stern-Gerlach magnet (SGM) is placed in front of the detector to deflect atoms in the $M=\pm1$ states. 
As the atoms travel from the optical pumping zone to the SGM, they pass through a probe region inside a magnetic shield. A cosine coil inside the shield generates a bias magnetic field. When an atom absorbs a photon from the probe laser, there is approximately a 1/3 chance that it will decay into the $M = 0$ state, which is then detected by a microchannel plate (MCP) detector located behind the SGM. The frequency of the probe laser is scanned to record the spectrum. Further details of the experimental setup and procedure can be found in Refs.~\cite{Chen2020PRA_He, Zheng2017PRL_He_Lamb}.

During our experimental evaluation, we performed a comprehensive analysis of the sources of uncertainty in our measurements. A detailed breakdown of these sources is provided in Table~\ref{errbar}. Our primary objective was to elucidate the key factors contributing to these uncertainties.

\begin{table}[hb]
\renewcommand{\arraystretch}{1.2}
\caption{Uncertainty budget of the $2^3S_1 - 2^3 P_0$ transition frequency $(f_0)$, in kilohertz.}		
\begin{tabular}{l c c}	
    \hline \hline
    Source & Corrections & $\Delta f (1 \sigma )$ \\
    \hline
    Statistics & & 0.22 \\
    First-order Doppler & & 0.82 \\
    Second-order Doppler & & 0.01 \\
    Frequency calibration & & 0.06 \\
    Quantum interference & 0.05 & 0.02 \\
    Laser power & & 0.20 \\
    Zeeman effect & & 0.01 \\
    Recoil shift & -42.48 & 0.01 \\
    Total & 276 764 094 712.45 & 0.86 \\
    \hline \hline		
\end{tabular}
\label{errbar}
\end{table}

To improve the probing method, we replaced the conventional standing wave with a traveling wave. After six months of measurements, this modification reduced the statistical uncertainty to approximately 0.22~kHz. Further details about the sequential counter-propagating traveling-wave optical probing (SCTOP) technique can be found in our earlier work\cite{Wen2023PRA_traveling}.

We assessed the residual first-order Doppler effect by measuring at different longitudinal velocities ranging from 100 to 400 m/s. The data were fitted with a linear model, yielding an intercept of approximately 0.82 kHz. The slope of the fit provided an estimate of the residual first-order Doppler shift, which was found to be less than 10~$\mu$rad, corresponding to 0.036(32)~kHz/(m/s).

The second-order Doppler shift was also calculated for each velocity. Given the beam velocity distribution of 5m/s, the maximum uncertainty due to this effect was about 0.01~kHz.

Our laser systems were phase-locked to a reference laser stabilized with an ultra-low expansion (ULE) cavity with a finesse of approximately 300,000. The frequency of the reference laser was determined by beating it against a frequency comb, which was referenced to a hydrogen maser clock with a stability better than $10^{-13}$. The upper limit of the calibration uncertainty was approximately 0.06~kHz.

The quantum interference shift, calculated using the same method as Marsman \textit{et al.}~\cite{Marsman2012PRA_QI}, was found to be +0.05(2)~kHz.

During the experiment, the laser power was stabilized and maintained below 1~$\mu W$. Power was measured at different velocities: 0.3, 0.6, and 0.9~$\mu W$ for velocities between 250 and 400~m/s, and 0.2, 0.4, and 0.6~$\mu W$ for velocities between 100 and 250~m/s. The final result for each velocity was obtained by extrapolating the frequencies to zero laser power. A Thorlabs PM200D detector, with a stated nonlinearity of 0.5\%, was used to measure the power. An appropriate filter was installed to mitigate interference from other laboratory lasers. The uncertainty from this effect was capped at less than 200~Hz.

The introduction of a magnetic shield in the probe zone reduced the Earth's magnetic field to below 10 micro-Gauss. Additionally, a cosine coil produced a controlled magnetic field to eliminate the first-order Zeeman shift. The center frequency of two Zeeman peaks was derived from this setup. The magnetic field strength, measured to be around 15.7~Gauss at an 8~A current, was used to correct for the second-order Zeeman shift, with an associated uncertainty below 0.01~kHz.

The recoil shift correction was calculated to be -42.48~kHz, using the formula $\Delta f_{recoil} = - h/(2m\lambda^2) = - ( h \nu^2 ) / ( 2 m c^2 )$.  The constants for this calculation were obtained from CODATA2018\cite{Tiesinga2021RevModPhys_CODATA2018}, and the associated uncertainty was less than 0.01~kHz.

After thoroughly reviewing all possible sources of uncertainty, no additional systematic errors were identified. As a result, the transition frequency for the $2^3 S_1 - 2^3 P_0$ transition was determined to be $f_0=276,764,094,712.73(86)$~kHz. This value is consistent with the result of Cancio Pastor \textit{et al.}\cite{Pastor2004_He4, Pastor2006_He4Err}, within two combined standard deviations.

\begin{table}[hb]
\renewcommand{\arraystretch}{1.2}
\caption{Comparison of centroid frequency $(f_c)$ of $2^3S - 2^3P$ transition of $^4$He, in kilohertz.}
\begin{tabular}{l l l}
    \hline \hline
    &$f_c - 276\,736\,495\,000$ & $\Delta f (1 \sigma ) $ \\
    \hline
    Shiner \textit{et al}.~\cite{Shiner1994PRL_He4} & 580 & 70 \\
    Cancio Pastor \textit{et al}.~\cite{Pastor2004_He4,Pastor2006_He4Err} & 649.5 & 2.1 \\
    Zheng \textit{et al}.~\cite{Zheng2017PRL_He_Lamb} & 600.0 & 1.6 \\
    Zheng \textit{et al} + Correction & 654.5& 2.1 \\
    \hline
    Patk{\'o}{\v{s}} \textit{et al}.~\cite{Patkos2021PRA_He_Lamb} & 620 & 54 \\
    \hline
    This work & 655.21 & 0.87\\
    \hline \hline
\end{tabular}
\label{centroid}
\end{table}

\begin{table*}[htb]
\renewcommand{\arraystretch}{1.2}
\caption{Calulation and comparison of the $\delta r^2 [^3\textrm{He}-^4\textrm{He}]$ values obtained from different methods}
\begin{tabular}{l r r }
    \hline \hline
    $\nu(^3\textrm{He}, 2^3 S - 2^3 P )$ (centroid) & 276\,702\,827\,204.8 (2.4)~kHz & Exp.~\cite{Pastor2012PRL_He3}  \\
    $-\delta E_{iso}$ (point nucleus)& +33\,667\,149.3 (0.9)~kHz & Theo.~\cite{Pachucki2017PRA_He_Theo}  \\
    $-\nu(^4\textrm{He}, 2^3 S - 2^3 P)$ (centroid)& -276\,736\,495\,655.21 (87)~kHz & this work  \\
    $\delta \nu_{FNS}$& -1\,245.9 (2.9)~kHz &  \\
    $C(2^3 S - 2^3 P)$ & -1\,212.2 (1) kHz/fm$^2$ & Theo.~\cite{Pachucki2017PRA_He_Theo}  \\
    $\delta r^2, (2^3 S - 2^3 P)$& 1.0733 (21)~fm$^2$   &  \\
    \hline
    $\delta r^2, (2^3S - 2^1 S)$& 1.0757 (15)~fm$^2$  & Exp.~\cite{Werf2023arxiv_he3_trap} \\
    $\delta r^2, \mu\textrm{-He}^+ (2S-2P)$& 1.0636 (31)~fm$^2$  & Exp.~\cite{CREMA2023arxiv_uhe3} \\
    \hline \hline
\end{tabular}
\label{is_cal}
\end{table*}

By combining this result with precise measurements of the intervals between the $2^3 P_J$ states~\cite{Zheng2017PRL_FS}, we were able to determine the centroid frequency for the $2^3S-2^3P$ transition. The reported centroid frequency, as shown in Table~\ref{centroid}, is $f_c=276\, 736\, 495\, 655.48(87)$~kHz. Notably, Zheng's result\cite{Zheng2017PRL_He_Lamb} agrees with our value after adjusting for the post-selection (PS) effect. However, it is important to re-evaluate the findings of Shiner's group~\cite{Shiner1994PRL_He4,Shiner1995PRL_He3}, as their results may also be affected by the PS effect.

By combining the centroid frequency of the $2^3S - 2^3P$ transition in $^4$He with the corresponding result for $^3$He provided by Cancio Pastor \textit{et al.}\cite{Pastor2012PRL_He3}, we derived a new value for the squared nuclear charge radius difference between $^3$He and $^4$He, $\delta r^2 [r_{h}^{2} - r_{\alpha}^{2}]$, as 1.074(2)~fm$^2$, based on the latest theoretical predictions\cite{Pachucki2017PRA_He_Theo}. This result shows excellent agreement with the recently updated value from the $2^3 S - 2^1 S$ transition measurements\cite{Rengelink2018_He4_trap,Werf2023arxiv_he3_trap}, but exhibits a significant discrepancy compared to the recent $\mu\textrm{-He}^+$ measurement~\cite{Krauth2021Nature_uHe4,CREMA2023arxiv_uhe3}.

%%%%%%%%%%%%%%%%%%%%%%%%%%%%%%%%%%%%%%%%%%%%%%%%%%%%%%%%%
\section{Simple phenomenologic model for the Post-selection Shift} ~\label{app:phenome}
%%%%%%%%%%%%%%%%%%%%%%%%%%%%%%%%%%%%%%%%%%%%%%%%%%%%%%%%%

The post-selection shift (PSS) could be described using a simple phenomenological model. As illustrated in Fig.~\ref{fig:pheno}, there are two slits located at both sides of the beam. To simplify the model, we assume that the slits are narrow enough to allow only one atom to pass through at a time. The atom interacts with the laser, and after spontaneous emission, its transverse velocity changes from $v_0$ to $v_1$. The velocity change due to the photon recoil, $v_{photon}$, can be expressed as:
\begin{equation}
    v_{photon} = v_R \cos{\theta} = \frac{\hbar k}{m} \cos{\theta}
\end{equation}
where $\cos{\theta}$ represents the projection of the spontaneous emission in the x direction. In the ideal case, we can set $\theta = 0$, so $v_{photon} = v_R$, where $v_R$ is the recoil velocity. Taking into account the geometric configuration, we get:
\begin{equation}
    \begin{split}
        v_0 \frac{\ell}{v} + v_1\frac{L}{v} = 0 \\
        v_1 = v_0 + v_R
    \end{split}    
\end{equation}
where $\ell$ is the distance from the atom-laser interaction point to the first slit, and $L$ is the distance to the second slit. This leads to a simplified expression for the PSS:
\begin{equation}
    \Delta \nu = \frac{\vec{k} \cdot \vec{v}}{2\pi } =  \frac{kv_0}{2\pi} = - \frac{\nu_c}{c} v_R \frac{L}{\ell+L} = - \frac{h}{mc^2}\frac{L}{\ell+L}\nu_c^2
\end{equation}
where $\nu_c$ is the detected transition frequency. For the experiment by Zheng \textit{et al.}~\cite{Zheng2017PRL_He_Lamb}, with $\ell = 0.55~m$ and $L = 1.05~m$, the estimated PSS is about $-55.7$~kHz. While this method does not provide a precise calculation of the PSS, it offers a reasonable estimate of its magnitude.

We can also estimate the impact of PSS on the measurement of the $2^3 P_J$ interval of helium using this simple model. For the larger interval, $\nu_{02} = 2^3 P_0 - 2^3 P_2$, the PSS-induced shift can be expressed as:
\begin{equation}
    \Delta \nu_{02} = - \frac{h}{mc^2}\frac{L}{\ell+L}(\nu_0^2-\nu_2^2)
\end{equation}
where $\nu_0$ and $\nu_2$ represent the frequencies of the $2^3S_1-2^3P_0$ and $2^3S_1-2^3P_2$ transitions, respectively. For the experiment reported by Zheng \textit{et al.}~\cite{Zheng2017PRL_FS}, this shift is around $-13$~Hz, which has a negligible impact on the overall results.

%%%%%%%%%%%%%%%%%%%%%%%%%%%%%%%%%%%%%%%%%%%%%%%%%%%%%%%%%
\section{Theory of post-selection effect}~\label{app:theory}
%%%%%%%%%%%%%%%%%%%%%%%%%%%%%%%%%%%%%%%%%%%%%%%%%%%%%%%%%
\subsection{Theoretical model}

\begin{figure*}[ht]
\centering
\subfloat{\label{fig:c1a}}
\subfloat{\label{fig:c1b}}
\includegraphics[width=0.75\textwidth]{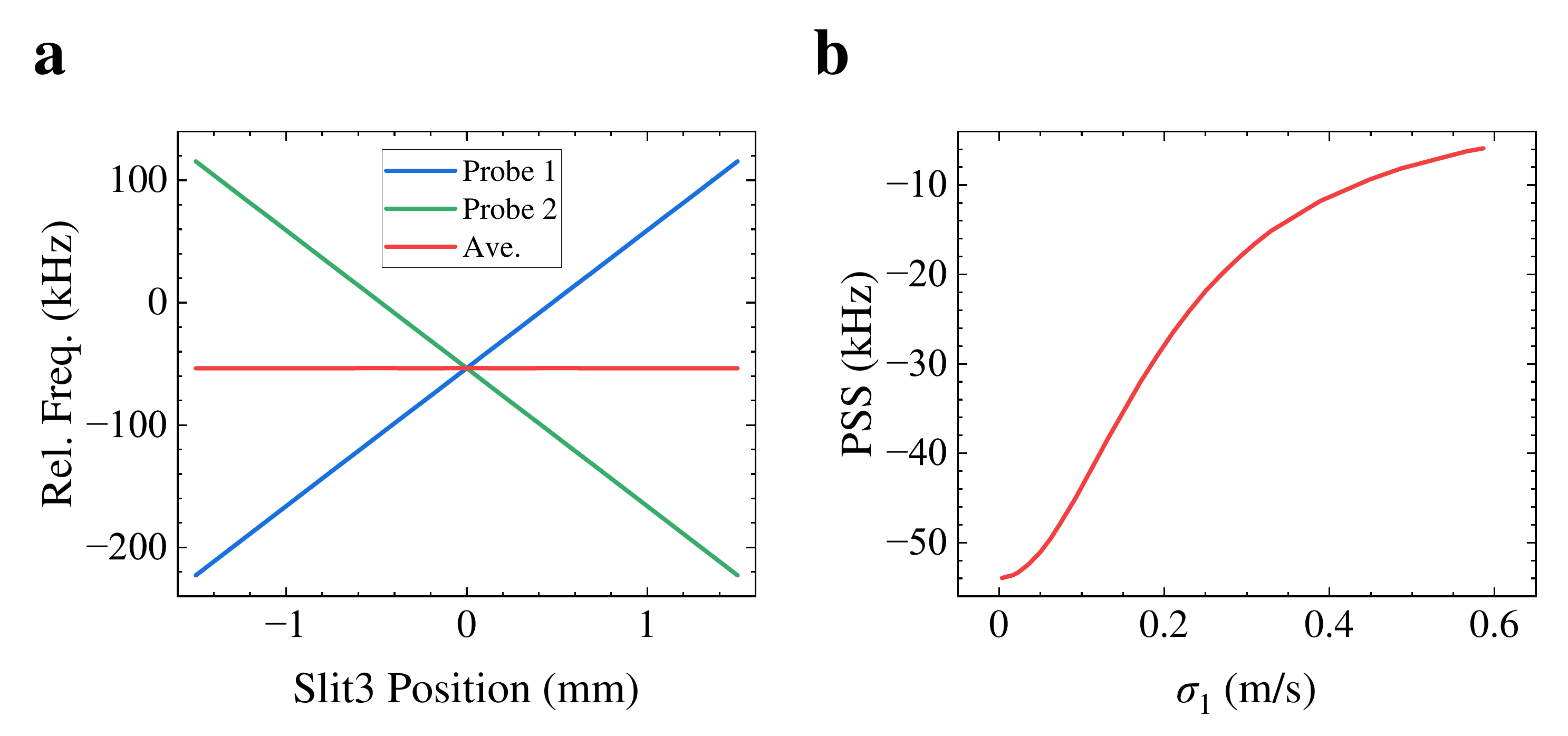}
\caption{
\textbf{a}, The center frequency $\delta$ verse center velocities $v_{1,b}$ of the detection ranges, which is determined by the position of slit3.
\textbf{b}, The average center frequency $\delta$ verse different range of detection $\sigma_{1}$, which is determined by the size of slit3.}
\label{fig:theory_result}
\end{figure*}

The experimental measurement process is comprised of three main steps: initial state preparation, coupling, and detection.
For the initial state preparation, we model the velocity distribution of the incident atomic beam in the x-direction as Gaussian, and the probability density of the initial velocity $v_{0}$ is given by
\begin{equation}
    P_{\mathrm{density}}(v_{0})=\frac{1}{\sqrt{2\pi}\sigma_{0}}e^{-\frac{(v_{0}-v_{0,b})^{2}}{2\sigma_{0}^{2}}},
\end{equation}
where $v_{0,b}$ represents the mean and $\sigma_{0}$ the standard deviation of the distribution.

During the coupling process, atoms traveling at a velocity $v_{0}$ can be excited with a probability defined by
\begin{equation}
    P_{exc}=\frac{\Omega^{2}}{(\delta-kv_{0})^{2}+\gamma^{2}},
\end{equation}
where $\Omega$ is the Rabi frequency associated with the coupling between the probe laser (frequency $\omega_{p}$) and the corresponding atomic transition (frequency $\omega_{12}$).
Here, $\delta=\omega_{p}-\omega_{12}$, $k$ is the wave vector of the probe laser,
and $\gamma$ denotes the decay rate of the excited atomic state.
Following excitation, the atom spontaneously emits a photon with a wave vector directed along the x-axis, given by $k\mathrm{cos}\theta$,
where $\theta$ is the angle between the photon wave vector and the x-axis.
This emission alters the atom's velocity along the x-axis from $v_{0}$ to $v_{1}=v_{0}+v_{R}-v_{R}\mathrm{cos}\theta$, with $v_{R}$ representing the recoil velocity.
Consequently, the probability function for the change in atom velocity due to photon absorption and spontaneous emission is $P_t(v_0,v_1)=\frac{\mathrm{sin}\theta}{2}=\frac {1}{2}\sqrt{1-(v_1-v_0-v_R)^2/v_R^2}$ for $0\leq|v_1-v_0|\leq 2v_R$. 
It is important to note that during the coupling process, due to the short interaction time and weak coupling strength, we can reasonably assume that only one excitation and subsequent spontaneous emission occur. 

In the detection step, the detecting efficiency (the collection efficiency through the last slit) $\eta$ for various $v_{1}$ is determined by the position and size of the last slit in front of the MCP detector (slit3).
Here we express the detection efficiency as follows:
\begin{align}
    \eta(v_{1})=&e^{-\frac{(v_{1}-v_{1,b})^{2}}{2\sigma_{1}^{2}}}\nonumber\\
    =&e^{-\frac{(v_{0}+v_{R}-v_{R}\mathrm{cos}\theta-v_{1,b})^{2}}{2\sigma_{1}^{2}}}.
\end{align}
where $v_{1,b}=\xi x$ and $\sigma_{1}$ are determined by the position $x$ and the size of slit3, respectively, and $\xi$ is the slope that depends on the time of flight of the atom before detection.

In this measurement, only excited helium atoms can be detected, meaning the number of atoms detected depends on both the probability of excitation and the detection efficiency.
Based on the above discussion, we can express this relationship as
\begin{align}
    N\propto&\int_{-\infty}^{\infty}P_{\mathrm{density}}(v_{0})P_{exc}\int_{0}^{\pi}\eta(v_{1})\frac{\mathrm{sin}\theta}{2}\mathrm{d}\theta\mathrm{d}v_{0}\nonumber\\
    =&\int_{-\infty}^{\infty}\frac{1}{\sqrt{2\pi}\sigma_{0}}e^{-\frac{(v_{0}-v_{0,b})^{2}}{2\sigma_{0}^{2}}}\frac{\Omega^{2}}{(\delta-kv_{0})^{2}+\gamma^{2}}\cdot\nonumber\\
    &\int_{0}^{\pi} e^{-\frac{(v_{0}+v_{R}-v_{R}\mathrm{cos}\theta-v_{1,b})^{2}}{2\sigma_{1}^{2}}}\frac{\mathrm{sin}\theta}{2}\mathrm{d}\theta\mathrm{d}v_{0}.
\label{eq:N_1}
\end{align}

For an extremely narrow slit3, where $\sigma_{1}\ll\sigma_{0}$, the detection efficiency can be approximated by a Dirac function $\eta(v_{1})\simeq\delta(v_{1}-v_{1,b})$,
indicating that only atoms with a velocity of $v_{1}=v_{0}+v_{R}-v_{R}\mathrm{cos}\theta=v_{1,b}$ can be detected.
Since the probability of spontaneous photons emission at each angle $\theta$ is constant, we can simplify $v_{1}$ as $v_{1}=v_{0}+v_{R}$.
Assuming $\eta(v_{1})\simeq\delta(v_{1}-v_{1,b})$ and $\sigma_{1}\ll\sigma_{0}$, 
we obtain:
\begin{align}
    N\propto&\int_{-\infty}^{\infty}\frac{1}{\sqrt{2\pi}\sigma_{0}}e^{-\frac{(v_{0}-v_{0,b})^{2}}{2\sigma_{0}^{2}}}\frac{\Omega^{2}}{(\delta-kv_{0})^{2}+\gamma^{2}}\cdot\nonumber\\
    &\delta(v_{0}+v_{R}-v_{1,b})\mathrm{d}v_{0}\nonumber\\
    \propto&\frac{\Omega^{2}}{[\delta-k(v_{1,b}-v_{R})]^{2}+\gamma^{2}}.
\end{align}
Clearly, the center is $\delta_{+}=k(v_{1,b}-v_{R})$.
If the incident direction of the probe light is reversed, the center will change to $\delta_{-}=-k(v_{R}+v_{1,b})$.
Therefore, the average results for the frequency center of ``Probe 1'' and ``Probe 2'' is $\delta_{\mathrm{PS}, Ave}=-kv_{R}$.

Based on Eq.~\ref{eq:N_1}, 
the center frequencies of the spectra with post-selection at each $x$ position for two probes of light traveling in opposite directions
are shown in Fig.~\ref{fig:c1a}, using parameters $\sigma_{0}=0.2$~m/s, $v_{0,b}=0$~m/s, $\sigma_{1}=0.01$~m/s, $\Omega/2\pi=0.3$~MHz, $\gamma/2\pi=1.6$~MHz, $v_{R}=0.1$~m/s, $k=2\pi / \lambda =5.8\times10^{6}$~$\mathrm{m}^{-1}$. 
As shown in Fig.~\ref{fig:c1a}, the average center frequency is approximately 54~kHz, which is close to $kv_{R}/2\pi=84$~kHz and is consistent with our analysis.
Moreover, as depicted in Fig.~\ref{fig:c1b}, with $\sigma_{1}$ increasing from 0.1~m/s to 0.6~m/s, the average center frequency decreases from 54~kHz to 0.
In other words, the effect of post-selection on center frequency measurements becomes less significant with an increase in the size of slit3, as predicted in our previous work~\cite{Zheng2019PRA_LFS}. 

\begin{figure*}[t]
\centering
\subfloat{\label{fig:c2a}}
\subfloat{\label{fig:c2b}}
\subfloat{\label{fig:c2c}}
\subfloat{\label{fig:c2d}}
\includegraphics[width=0.85\textwidth]{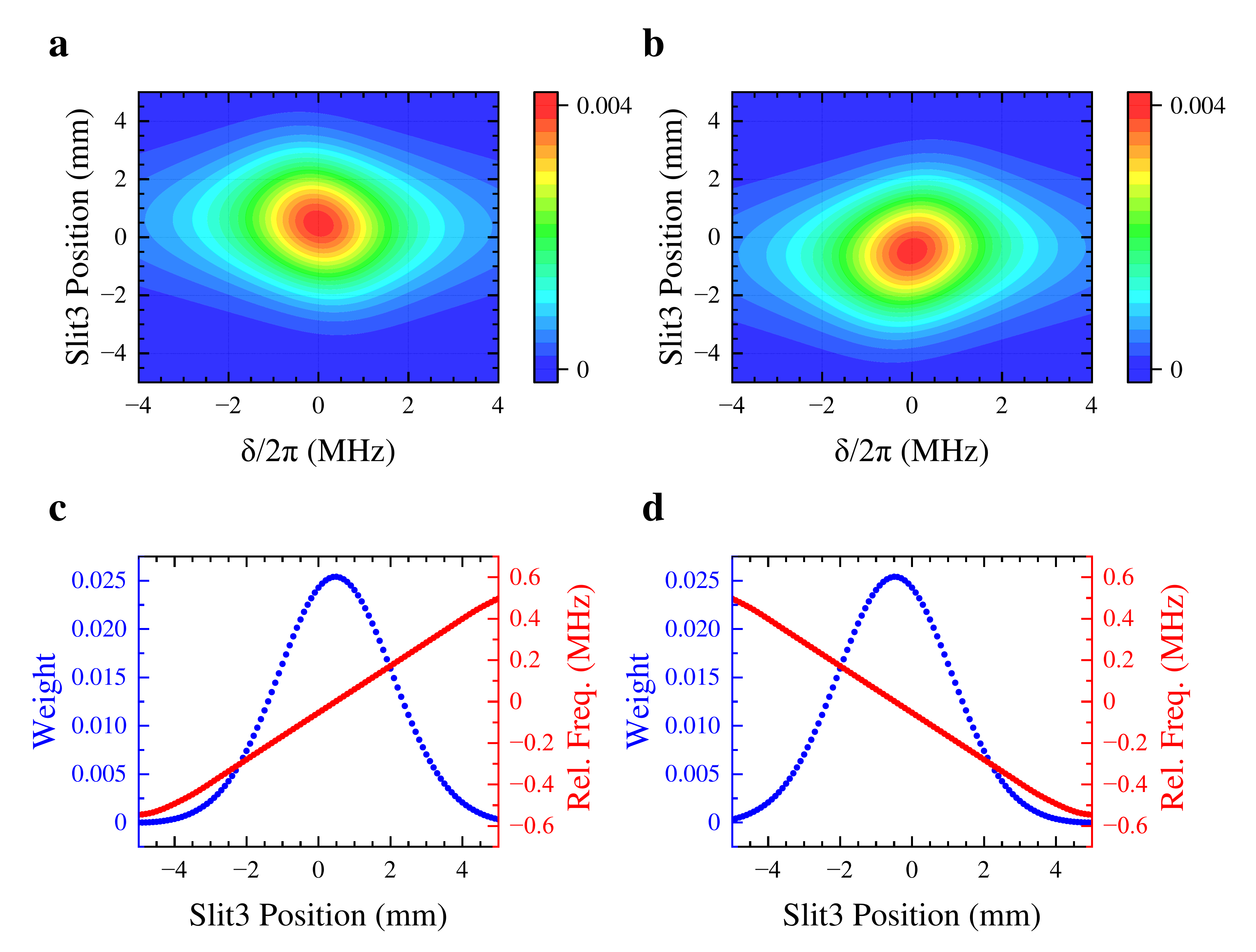}
\caption{
\textbf{Numerical results from the theoretical model.} \textbf{a} \& \textbf{b}, Theoretical results corresponding to Figs.~\ref{fig:2a} \&~\ref{fig:2c} in the main text respectively. \textbf{c} \& \textbf{d}, The peak value of the collection probability $P_{peak}$ (Weight) and the corresponding relative center frequencies of spectra obtained at various positions of slit3.}
\label{fig:theory_result_Fig2}
\end{figure*}

In the absence of slit3, and consequently without post-selection, 
the detection efficiency for the coupled atomic velocities is 1.
Thus, the number of detected atoms is determined solely by the probability of excitation, expressed as:
\begin{align}
    N\propto\int_{-\infty}^{\infty}\frac{1}{\sqrt{2\pi}\sigma_{0}}e^{-\frac{(v_{0}-v_{0,b})^{2}}{\sigma_{0}^{2}}}\frac{\Omega^{2}}{(\delta-kv_{0})^{2}+\gamma^{2}}\mathrm{d}v_{0}.
\end{align}

The center frequency is $kv_{0,b}$.
For the counter-propagating probe, the center frequency should be $-kv_{0,b}$.
Consequently, the average center frequency is 0, differing from the center frequency with post-selection.
The difference is influenced by the size of slit3.
If the slit3 is small enough to permit only one specific velocity to pass through ($\eta(v_{1})=\delta(v_{1}-v_{1,b})$), the value of PSS reaches the largest $kv_{R}$.
The PSS decreases as the size of slit3 increases, tending toward 0 when $\sigma_{1}\gg kv_{R}$, as shown in Fig.~\ref{fig:c1b}. 

%%%%%%%%%%%%%%%%%%%%%%%%%%%%%%%%%%%%%%%%%%%%%%%%%%%%%%%%%%%%%
\subsection{Numerical calculation} 

Based on Eq.~\ref{eq:N_1}, to calculate the number of detectable atoms, the parameters of $v_{0,b},\sigma_{0},\Omega,\gamma,v_{1,b},\sigma_{1},k$ and $v_{R}$ must be specified according to the experimental conditions. 
For the selected excited state of $^{4}$He, we have $\gamma/2\pi=1.62$~MHz.
The power of the probe light is $P = 1 \mu W$ with a diameter of 1~mm,
resulting in the Rabi frequency of $\Omega=2.1$~MHz and a recoil velocity of $v_{R}=\hbar k/m=0.09$~m/s, where $m$ is the mass of $^{4}$He.

\begin{figure*}[hbt]
\centering
\subfloat{\label{fig:d3a}}
\subfloat{\label{fig:d3b}}
\includegraphics[width=0.85\textwidth]{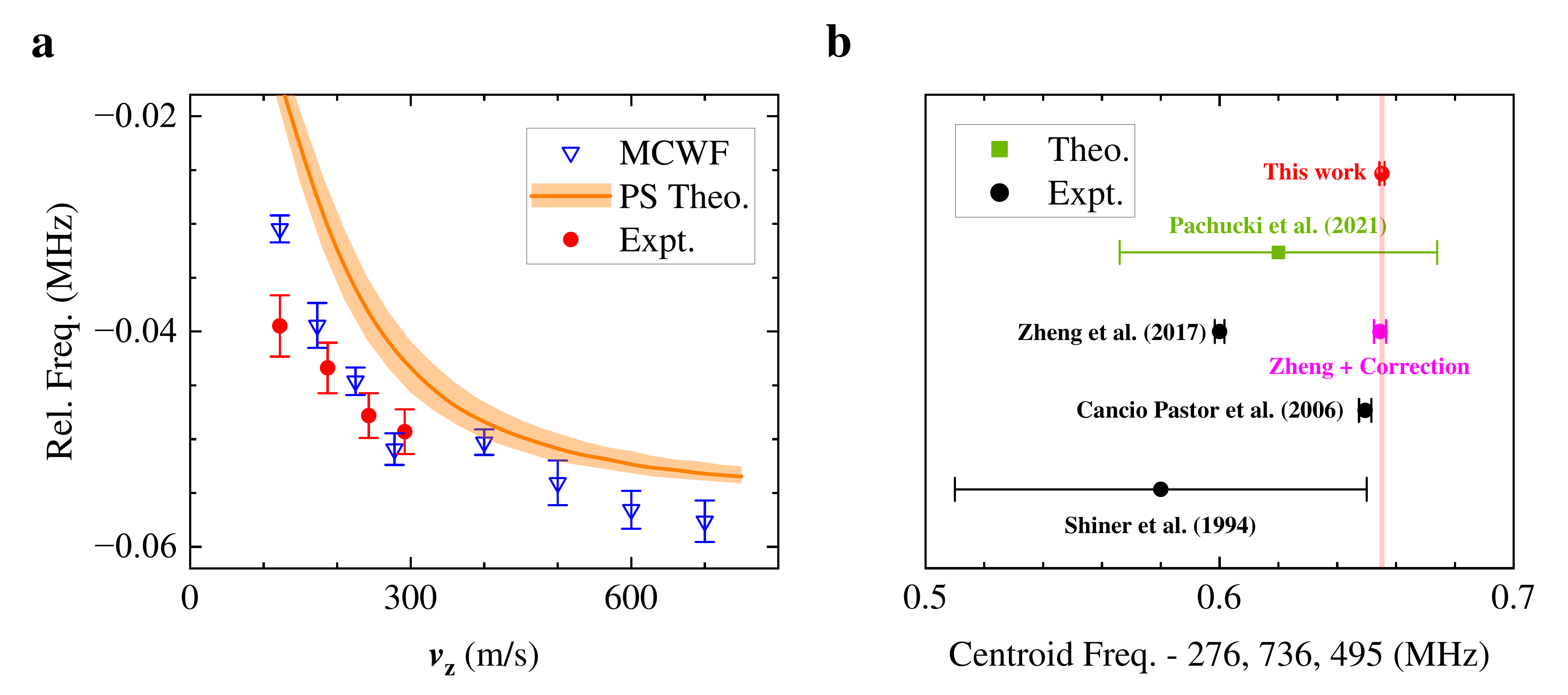}
\caption{\textbf{Post-selection shifts and correction.} \textbf{a}, The experimental and numerical results of PSS for various longitudinal atomic beam velocity $v_z$.
The error bars of the red and blue points represent statistical uncertainties, while the orange bands show the results with a variation of atomic initial lateral velocity distribution $\sigma_{0}$ by $\pm$10\%.
\textbf{b}, Comparison for the centroid frequency of the $2^3S - 2^3 P$ transition of $^4$He. The corrected results for Zheng \textit{et al.}~\cite{Zheng2017PRL_He_Lamb} using PSS has been included. The present result uses the fine-structure intervals from Zheng~\textit{et al.}~\cite{Zheng2017PRL_FS}.}
\label{fig:result}
\end{figure*}

In our experiment, the velocity distribution of the incident atomic beam is primarily determined by the two slits (shown on the left side of Fig.~\ref{fig:setup}, labeled slit1 and slit2) that constrain the atomic beam, particularly, the position ($x_{2},y_{2},z_{2}$) and diameter $\Delta x_{2}$ of the second slit (slit2).
Although the distance $l_{12}$ between slit1 and slit2 and the size of slit2 are fixed, $\sigma_{0}$ of the initial velocity distribution is influenced by the atomic longitudinal velocity $v_{z}$, as $\sigma_{0}\propto0.5\Delta x_{2}v_{z}/l_{12}$.
In our experiments, the velocity distribution of the incident atomic beam is characterized by $v_{0,b}=0$~m/s and $\sigma_{0}=0.1$~m/s with $v_{z}=290$~m/s. Consequently, $\sigma_{0}$ for different longitudinal velocities $v_z$ can be expressed as $\sigma_{0}(v_{z})=v_{z}\times0.1/290$ for the situation depicted in Fig.~\ref{fig:d3a} . However, for the situation shown in Fig.~\ref{fig:d3b} where slit2 is absent, the $\sigma_{0}$ is independent of $v_{z}$.

%%%%%%%%%%%%
%%%%%%%%%%%%

The main parameters $v_{1,b},\sigma_{1}$ of the detected efficiency $\eta$ are intuitively related to the position $x_{3},y_{3},z_{3}$ and diameter $\Delta x_{3}$ of the slit3, as well as influenced by the distance $\ell$ between slit2 and probe, the distance $L$ between probe and slit3, and the longitudinal velocities $v_z$.
For the case where $x_{3}=0,y_{3}=0$, only atoms with velocities $v_{1}$ that satisfy the conditions $-\Delta x_{3}/2<v_{0}t_{1}+v_{1}t_{2}<\Delta x_{3}/2$ can be detected, where $t_{1}=\ell/v_{z},t_{2}=L/v_{z}$.
Therefore, the center of $\eta$ is given by $v_{1,b}=v_{R}t_{1}/(t_{1}+t_{2})$ and $\sigma_{1}=\Delta x_{3}/[2(t_{1}+t_{2})]$.

By substituting the above parameters $v_{0,b},\sigma_{0},\Omega,\gamma,v_{1,b},\sigma_{1},k$ and $v_{R}$ into Eq.~\ref{eq:N_1}, we can derive the theoretical results presented in Figs.~\ref{fig:theory_result_Fig2}\&\ref{fig:result}.

Fig.~\ref{fig:theory_result_Fig2} illustrates the theoretical results corresponding to Fig.~\ref{fig:2a} \&~\ref{fig:2c} in the main text. From the results of Figs.~\ref{fig:c2a} \&~\ref{fig:c2b}, we can extract the peak collection probability $P_{peak}$ and the corresponding relative center frequencies for various detection positions (slit positions).
We define the weight $c_{j}$ as the ratio of the peak detection probability at each position to the total atomic excitation probability, given by $c_{j}=P_{peak,j}/\sum_{i}P_{peak,i}$, where $j$ denotes the $j-th$ detection position. consequently, the effective resonance frequency can be approximated as
\begin{equation}
    \omega_{eff} =\frac{\sum_{j}c_{j}^{\mathrm{probe1}}\omega_{\mathrm{peak},j}^{\mathrm{probe1}}+\sum_{j}c_{j}^{\mathrm{probe2}}\omega_{\mathrm{peak},j}^{\mathrm{probe2}}}{2}.
\label{eq:weff}
\end{equation}
Analyzing the data from Figs.~\ref{fig:c2c} \&~\ref{fig:c2d}, we obtain $\omega_{\mathrm{eff}}\simeq-0.43$~kHz. This effective frequency $\omega_{eff}$ closely approximates the transition frequency. 
However, since this is a mathematical approximation, the estimated value of $\omega_{\mathrm{eff}}$ is affected by the range and step size of the slit position chosen.

%%%%%%%%%%%%%%%%%%%%%%%%%%%%%%%%%%%%%%%%%%%%%%%%%%%%%%%%%
\section{Estimates of PSS in certain experimental configurations}~\label{app:pss}
%%%%%%%%%%%%%%%%%%%%%%%%%%%%%%%%%%%%%%%%%%%%%%%%%%%%%%%%%

Both the experimental results (Fig.~\ref{fig:2f}) and the theoretical analysis reveal a systematic frequency shift in Stern-Gerlach-type atomic beam spectroscopy measurements, attributed to post-selection effects. This shift, which we will refer to as the post-selection shift (PSS), was found to be $\Omega_{\mathrm{PSS}}\approx$ -55~kHz, slightly below the predicted amplitude of $kv_R=$ 84~kHz. This discrepancy can be explained by the practical limitation of the slit width $\Delta x_1$ in front of the detector, in contrast to the ideal scenario discussed above, where $\sigma_1\rightarrow 0$. 
The slit width $\Delta x_1$ determines the post-selection parameter, specifically $\sigma_1$, which is proportional to $\Delta x_1 v_z$. It is expected that the observed results will converge to the value without PSS if $\sigma_1\rightarrow\infty$. 

The PSS was further investigated by repeating the measurements with different longitudinal velocities, $v_z$, and the results are presented in Fig.~\ref{fig:d3a}. In particular, we observed that the PSS becomes more pronounced as $v_z$ increases, which is consistent with our prediction that $\sigma_0$ increases with $v_z$, and the PSS will eventually saturate at $kv_R$. 
To further analyze this effect, we employed numerical calculations based on more rigorous models, including the Monte Carlo Wave Function (MCWF) method and the full model described by Eq.~\ref{eq:dist}, incorporating practical experimental parameters. A detailed description of these numerical methods can be found in the Methods section. 
As shown in Fig.~\ref{fig:d3a}, both models confirm the observed PSS effect, showing excellent agreement with the experimental data.
The red theory line deviates slightly from the experimental data because it represents an approximate analytical solution rather than a specific numerical solution.

Our results highlight the importance of carefully accounting for PSS in precise frequency measurements of the $2^3 S-2^3 P$ transition in helium atoms. 
In our experiments, we were able to eliminate the systematic frequency shift caused by PSS by removing the slit in front of the detector, ensuring that the detector collects all $M=0$ atoms after the Stern-Gerlach magnet.
This was achieved using the SCTOP (sequential counter-propagating traveling-wave optical probing) method~\cite{Wen2023PRA_traveling} and a narrow incident atomic beam.  
With the modified experimental setup, we measured the $2^3 S_1-2^3 P_0$ transition frequency of $^4$He to be $f_0 = 276\,764\,094\,712.45 \pm 0.86$~kHz. By combining the fine structure intervals presented in a previous study~\cite{Zheng2017PRL_FS}, we determined the centroid frequency of  $2^3 S-2 ^3 P$ to be $f_c = 276\,736\,495\,655.21 \pm 0.87$~kHz, as indicated by the red circle and belt in Fig.~\ref{fig:d3b}. 

In certain experimental configurations, the PSS can be derived and corrected. For example, the experimental by Zheng \textit{et al.}~\cite{Zheng2017PRL_He_Lamb}, the PSS is estimated to be around -55~kHz. The corrected transition frequency is depicted as a pink dot in Fig.~\ref{fig:d3b}, which agrees well with our measurements. 

For comparison, the frequencies measured in other studies are also plotted in Fig.~\ref{fig:d3b}. 
The $2^3 S - 2^3 P $ centroid frequency difference between this work and that of Cancio Pastor \textit{et al.}~\cite{Pastor2004_He4,Pastor2006_He4Err} is 5.7~kHz, about 2.5 times the combined uncertainty.
The value obtained by Shiner's group~\cite{Shiner1994PRL_He4} is consistent with our result, though with a larger uncertainty and also affected by the post-selection effect, which should be corrected. 

Additionally, Pachucki \textit{et al.} have completed the $\alpha^7m$ QED calculation for this transition with an accuracy of 54~kHz~\cite{Patkos2021PRA_He_Lamb}, and their result is consistent with the experimental data presented here. However, the theoretical precision is not yet sufficient to determine the helium charge radius directly from the $^4$He transition alone.

%%%%%%%%%%%%%%%%%%%%%%%%%%%%%%%%%%%%%%%%%%%%%%%%%%%%
\section{Monte Carlo Wave Function Method}~\label{app:MCWF}
%%%%%%%%%%%%%%%%%%%%%%%%%%%%%%%%%%%%%%%%%%%%%%%%%%%%

We use the Monte Carlo Wave Function (MCWF) method to analyze the evolution of atoms in the laser field, which is equivalent to the density matrix method. MCWF requires less computation ($\propto N$) than the density matrix method ($\propto N^2$).

We consider a three-level atomic system consisting of the excited state $\ket{e}$, the ground state $\ket{g}$, and the dark state $\ket{0}$. Initially, these atoms are all transferred to the ground state $\ket{g}$ by the pump light. In our experiment, due to the Stern-Gerlach magnet, only those atoms that end up in the dark state $\ket{0}$ contribute to the detection signal. The direction of the atomic beam is denoted by the z-axis, while the probe light propagates along the x-axis and the direction of the magnetic field is the y-axis. For simplicity, we assume that the longitudinal velocity $v_z$ of the atomic beam along the z-axis is uniform. For the transverse velocity, we consider only he velocity $v_x$ in the x-axis direction.

Suppose that at time $t$ an atom is in a state with a normalized wave function of $\ket{\Psi (t)}$. First, we consider the case where no quantum jump occurs. After a time $\delta t$, the evolution of the system is $\ket {\Psi ^{(1)} (t + \delta t)} $. The effective Hamiltonian of the evolution can be written as
\begin{equation}
\label{eq:Heff1}
    H_{eff} =\frac{ \vec{P}^2}{2m} + \hbar \omega_0 \ket{e}\bra{e} - \vec{d}\cdot\vec{E} - \frac{i \hbar}{2}\sum_{m}{C_m^{\dag} C_m}
\end{equation}
where $C_m$ is the quantum jump operator, there
\begin{equation}
\label{eq:Cm}
    \sum_{m}{C_m^{\dag} C_m = \Gamma \ket{e}\bra{e}}.
\end{equation} 
After a sufficiently small time $\delta t$, at the time $t + \delta t$, the wave function of the system evolves to
\begin{equation}
\label{eq:psi1}
    \ket{\Psi^{(1)} (t + \delta t)} = (1 - \frac{i H_{eff} \delta t}{\hbar}) \ket{\Psi (t)}.
\end{equation}
Here we introduce $\delta p$, which has the form
\begin{equation}
\label{eq:Jump1}
    \delta p = \delta t \frac{i}{\hbar}\bra{\Psi(t)}H_{eff} - H_{eff}^{\dag}\ket{\Psi(t)} = \sum_{m}{\delta p_m},
\end{equation}
\begin{equation}
\label{eq:Jump2}
    \delta p_m =\delta t \frac{i}{\hbar}\bra{\Psi(t)}C_m^{\dag} C_m\ket{\Psi(t)} \geq 0 .
\end{equation}

\begin{figure*}[ht]
\centering
\subfloat{\label{fig:e4a}}
\subfloat{\label{fig:e4b}}
\subfloat{\label{fig:e4c}}
\subfloat{\label{fig:e4d}}
\includegraphics[width=0.85\textwidth]{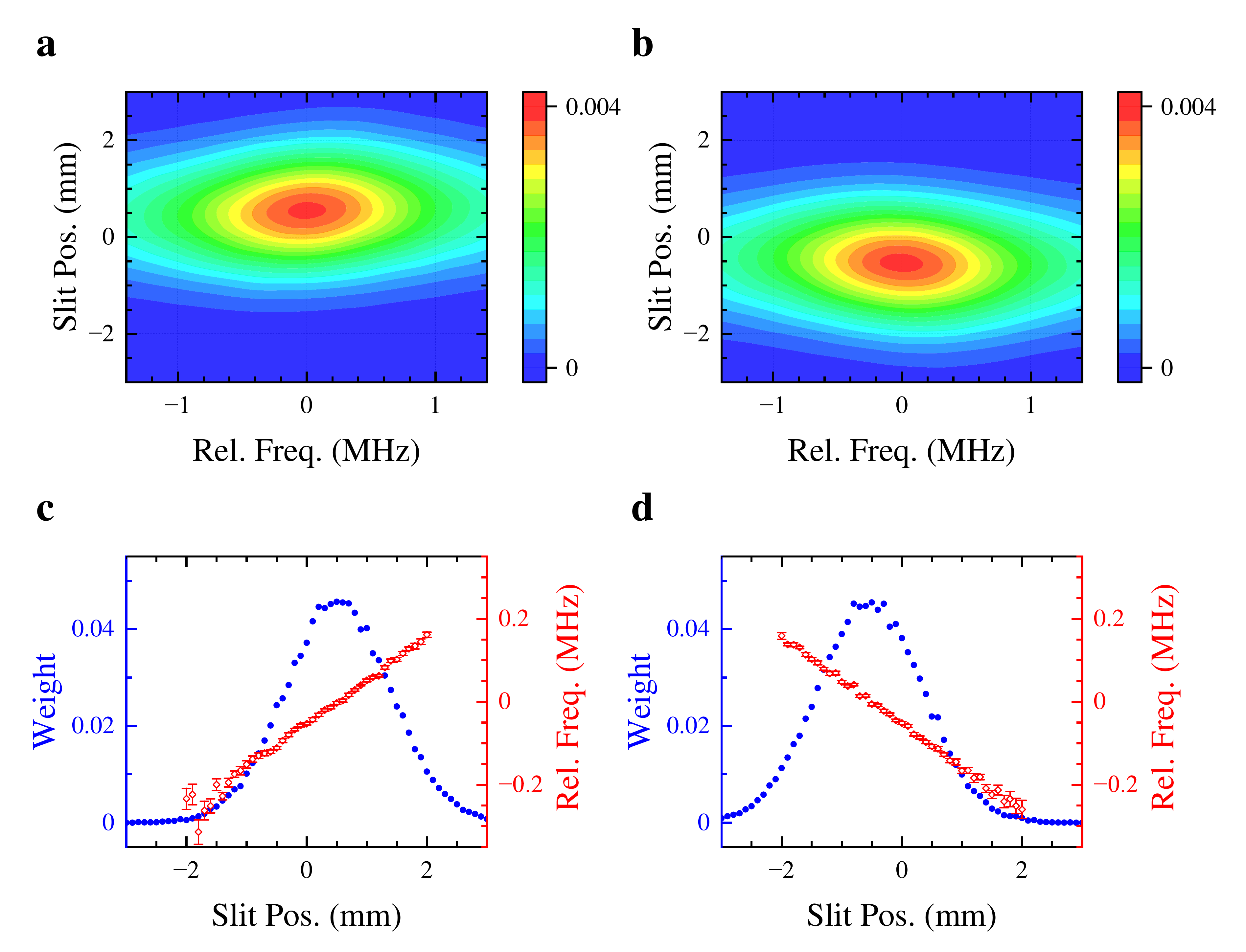}
\caption{
\textbf{Results obtained from the MCWF method.} \textbf{a} \& \textbf{b}, Correspond to Figs.~\ref{fig:2a} \&~\ref{fig:2c} in the main text, respectively. \textbf{c} \& \textbf{d}, Relative center frequencies and the corresponding weights for the spectra obtained at different positions of slit3.}
\label{fig:theory_result_Fig4}
\end{figure*}

When we consider a quantum jump, $\delta p$ is the probability of a quantum jump occurring, and $\delta p_m$ is the probability corresponding to each ``direction'' of the jump. We can use a random number $\epsilon$ uniformly distributed over $[0-1]$ to compare the size with $\delta p$ to determine whether a quantum jump has occurred. If $\epsilon > \delta p$, then no quantum jump occurs, and we normalize $\ket{\Psi^{(1)} (t + \delta t)}$ to get the wave function of the system at time $t + \delta t$.
If $\epsilon < \delta p$, then a quantum jump occurs, and the probability of $m$ possible quantum jumps is $\delta p_m/\delta p$. The wave function of the $t + \delta t$ time system is
\begin{equation}
\label{eq:psi3}
    \ket{\Psi (t + \delta t)} = \frac{C_m \ket{\Psi (t)}}{\Vert C_m \ket{\Psi (t)} \Vert} = \frac{C_m \ket{\Psi (t)}}{\sqrt{\delta p_m / \delta t}}.
\end{equation}

For the system we are studing, we can use the base vector of $\{\ket{e,p}, \ket{g,p} \}$. The momentum space is discretized in steps of $\hbar k$, where $k$ is the wave vector of light. Therefore, the normalized wave function can be written as
\begin{equation}
\label{eq:psi4}
    \ket{\Psi (t)} = \sum_{n}{c_{e,n}(t)\ket{e, n \hbar k} + c_{g,n}(t)\ket{g, n \hbar k}},
\end{equation}
where the coefficients $c_{e,n}(t)$ and $c_{g,n}(t)$ correspond to the probability amplitude at time $t$ of the atom in the excited state and the ground state, respectively, and the momentum projection along the x-axis is $n \hbar k$.

When an atom absorbs a photon from the laser field, it transitions from $\ket{g, n \hbar k}$ to $\ket{e, (n+1) \hbar k}$ or $\ket{e, (n-1) \hbar k}$, positive or negative depending on whether the light propagates forward or backward along the x-axis. 
When an atom is stimulated by the laser field to emit a photon, it transitions from $\ket{e, n \hbar k}$ to $\ket{g, (n+1) \hbar k}$ or $\ket{g, (n-1) \hbar k}$, positive or negative as the emitted photon propagates forward or backward along the x-axis. 
When an atom spontaneously emits a photon, to discretely consider the velocity change in the x-direction caused by the recoil momentum of the photon on the atom, we consider the projection of the recoil momentum of the photon on the atom along the x-direction as the transverse initial velocity change of the atom along the x-axis. 
After the spontaneous radiation, the atom transits from $\ket{e, n \hbar k}$ to $\ket{g, n \hbar k}$, and the initial velocity along the x-direction becomes
\begin{equation}
\label{eq:vx_sp}
    v_{x} ^{(1)} = v_x - \frac{\hbar k}{m} \sin \theta \cos \phi,
\end{equation}
where $\theta$ and $\phi$ are directions of the spontaneously radiated photon in the solid angle of space. At this point, we update the initial speed to $v_x = v_{x} ^{(1)}$.

If the atom spontaneously radiates into the dark state $\ket{0}$, where it no longer interacts with light, the wave function probability amplitude of the atom remains and the atom leaves the laser field.

\begin{equation}
\label{eq:cen2}
    c_{e,n} = 0
\end{equation}

\begin{equation}
\label{eq:cgn2}
    c_{0,n}= \frac{c_{e,n} ^{(1)} (t + \delta t) }{\sum_{m = - \infty} ^{+ \infty} {| c_{e,m} ^{(1)} (t + \delta t)|^2}}
\end{equation}

If the atom spontaneously radiates to the $m=-1$ state, the atom will not interact with the light field. But atoms in this state cannot reach the detector placed after the Stern-Gerlach magnet, so we no longer consider them. If the atom spontaneously radiates into the $\ket{g}$ state, then the normalized wave function probability of the atom at $t + \delta t$ is
\begin{equation}
\label{eq:cen3}
    c_{e,n} (t + \delta t) = 0,
\end{equation}
\begin{equation}
\label{eq:cgn3}
    c_{g,n}(t + \delta t) = \frac{c_{e,n} ^{(1)} (t + \delta t) }{\sum_{m = - \infty} ^{+ \infty} {| c_{e,m} ^{(1)} (t + \delta t)|^2}}.
\end{equation}

After obtaining the normalized wave function probability amplitude of the atom at the time of $t + \delta t$, the above process is repeated until the atom leaves the laser field, and we obtain the final wave function probability amplitude.		

Due to the presence of the Stern-Gerlach magnet, only atoms on the $\ket{0}$ state can be detected by the MCP detector. For these atoms, we can get their transverse velocity after passing through the probe light field, which is
\begin{equation}
\label{eq:vxn}
    v_{x,n} = v_x + \frac{n \hbar k}{m}.
\end{equation}
The corresponding probabilities are $| c_{0,n} |^2$.

By taking into account the longitudinal velocity of the atom, we can calculate the motion time of the atom before and after the probe light, and then obtain the transverse velocities $v_x$ and $v_{x,n}$ before and after passing through the probing region, respectively. Finally, we derive the transverse position of the atom at the slit in front of the detector, to determine whether the atom can be detected by the detector through the slit.

Using the MCWF method, we can simulate the theoretical results corresponding to Figs.~\ref{fig:2a} \&~\ref{fig:2c} in the main text. Based on the results of Figs.~\ref{fig:e4a} \&~\ref{fig:e4b}, we can obtain the relative center frequencies corresponding to different slit positions and the signal strength of the collected number of atoms. The results of center frequencies are weighted with the corresponding signal strengths, as shown in Figs.~\ref{fig:e4c} \&~\ref{fig:e4d}. According to Eq.~\ref{eq:weff}, we have $\omega_{\mathrm{eff}}/2\pi = 0.62$~kHz, which is within the numerical uncertainty. If the slit is placed at the lateral center position, the corresponding $\omega_{\mathrm{eff}}/2\pi = -52.16$~kHz have a significant systematic shift caused by the post-selection effect.		

\end{appendices}

%%%%%%%%%%%%%%%%%%%%%%%%%%%%%%%%%%%%%%%
\end{CJK*}

%\bibliographystyle{apsrev4-2}
%\bibliography{Bib_Slit}

\end{document}